\documentclass[a4paper,twocolumn,noarxiv,11pt,accepted=2022-03-16]{quantumarticle}
\pdfoutput=1
\usepackage{amsmath}
\usepackage[numbers,sort&compress]{natbib}
\usepackage{mathrsfs}
\usepackage{cases}
\usepackage{mathtools}
\DeclarePairedDelimiter\ceil{\lceil}{\rceil}

\usepackage[table,xcdraw]{xcolor}
\usepackage{hyperref}
\usepackage{todonotes}
\usepackage{qcircuit}
\usepackage{graphicx}
\usepackage[T1]{fontenc} 
\usepackage[utf8]{inputenc} 
\usepackage{amsfonts} 
\usepackage{bm}
\usepackage{physics}
\usepackage{verbatim}

\usepackage{amsthm}

\usepackage{enumitem}
\usepackage{multirow}
\usepackage{tabularx}
\usepackage{longtable}
\usepackage{float}
\usepackage{soul}

\newcommand{\vertiii}[1]{{\left\vert\kern-0.25ex\left\vert\kern-0.25ex\left\vert #1 
    \right\vert\kern-0.25ex\right\vert\kern-0.25ex\right\vert}}

\usepackage{hyperref}
\usepackage[capitalise,nameinlink]{cleveref}
\crefname{ansatz}{Ansatz.}{Ansatzes.}

\begin{document}

\title{Hybrid quantum-classical approach for coupled-cluster Green's function theory}

\author{Trevor Keen} 
\email{tkeen1@vols.utk.edu} 
\address{Department of Physics and Astronomy, University of Tennessee, Knoxville, Tennessee 37996, United States of America}
\author{Bo Peng}
\email{peng398@pnnl.gov} 
\address{Physical Sciences and Computational Division, Pacific Northwest National Laboratory, Richland, Washington 99354, United States of America}
\author{Karol Kowalski}
\email{karol.kowalski@pnnl.gov} 
\address{Physical Sciences and Computational Division, Pacific Northwest National Laboratory, Richland, Washington 99354, United States of America}
\author{Pavel Lougovski}
\thanks{Presently at Amazon Web Services}
\email{plougovski@gmail.com}
\address{Quantum Information Science Group, Computational Sciences and Engineering Division, Oak Ridge National Laboratory, Oak Ridge, Tennessee 37831, United States of America}
\author{Steven Johnston}
\email{sjohn145@utk.edu} 
\address{Department of Physics and Astronomy, University of Tennessee, Knoxville, Tennessee 37996, United States of America}
\address{Institute for Advanced Materials and Manufacturing, University of Tennessee, Knoxville,  Tennessee 37996, United States of America}
\date{\today}

\begin{abstract}
The three key elements of a quantum simulation are state preparation, time evolution, and measurement. While the complexity scaling of time evolution and measurements are well known, many state preparation methods are strongly system-dependent and require prior knowledge of the system's eigenvalue spectrum. Here, we report on a quantum-classical implementation of the coupled-cluster Green's function (CCGF) method, which replaces explicit ground state preparation with the task of applying unitary operators to a simple product state.
While our approach is broadly applicable to many models, we demonstrate it here for the Anderson impurity model (AIM). The method requires a number of $T$ gates that grows as $ \mathcal{O} \left(N^5 \right)$ per time step to calculate the impurity Green's function in the time domain, where $N$ is the total number of energy levels in the AIM. Since the number of $T$ gates is analogous to the computational time complexity of a classical simulation, we achieve an order of magnitude improvement over a classical CCGF calculation of the same order, which requires $ \mathcal{O} \left(N^6 \right)$ computational resources per time step.
\end{abstract}

\maketitle

\section{Introduction} 
Quantum simulations of fermionic and bosonic systems frequently involve three key components: state preparation, evaluating the system's dynamics, and measuring the relevant observables. While the computational complexity (e.g., the number of $T$ gates) of implementing dynamics and measurements on a quantum computer scale polynomially with the size of the system, the complexity of state preparation often depends strongly on the nature of the system and its ground state. Current quantum algorithms for computing zero temperature Green's functions, for instance, need to prepare the many-body ground state. Non-variational quantum algorithms that utilize adiabatic quantum computing and quantum phase estimation (QPE)~\cite{Bauer2016}, as well as variational approaches~\cite{Wecker2015, Peruzzo2014,PhysRevResearch.2.033281}, have been proposed. Their complexity depends on the size the spectral gap $\Delta$ and (or) the overlap between the ground state and the initial trial state $\gamma$, which implies prior knowledge of the system. In terms of resources, a system's ground state can be prepared to fidelity $1-\epsilon$ with probability $1-\vartheta$ using\cite{Lin2020nearoptimalground} 
\begin{equation*}
 \mathcal{O} \left( \frac{m\alpha}{\gamma \Delta} \left[ \log \frac{\alpha}{\Delta} \log\frac{1}{\gamma} \log \frac{\log\alpha/\Delta}{\vartheta} + \log\frac{1}{\epsilon} \right] \right)
\end{equation*}
single- and two-qubit gates, \emph{not} including query complexity of the block encodings of the Hamiltonian and trial state preparation. Here, $\alpha$ and $m$ describe the block encoding of the Hamiltonian of the system and $\gamma$ is the lower bound on the overlap of the initial trial state with the ground state. If a many-body system has a small spectral gap or the overlap $\gamma$ is exponentially small, initial state preparation may become infeasible on a quantum computer.

Here, we introduce a unitary formulation of the classical coupled-cluster (CC) method that can be implemented using hybrid quantum/classical computers to compute Green's functions. In contrast to other reported quantum-based Green's function approaches\cite{Kosugi2020, Endo2020}, our method does not employ the commonly-used variational quantum eigensolver (VQE). Instead, it replaces the need for preparing a many-body ground state on a quantum computer with a simpler task of applying unitary operators to a product state. This new approach thus eliminates the need for prior information about the system. While our approach is versatile and can be applied to many model Hamiltonians, we benchmark it here for the case of the Anderson Impurity model (AIM), which lies at the heart of other widely-used many-body approaches like dynamical mean-field theory (DMFT)~\cite{Georges1992}.

\begin{figure}
    \centering
    \includegraphics[width=0.85\columnwidth, keepaspectratio]{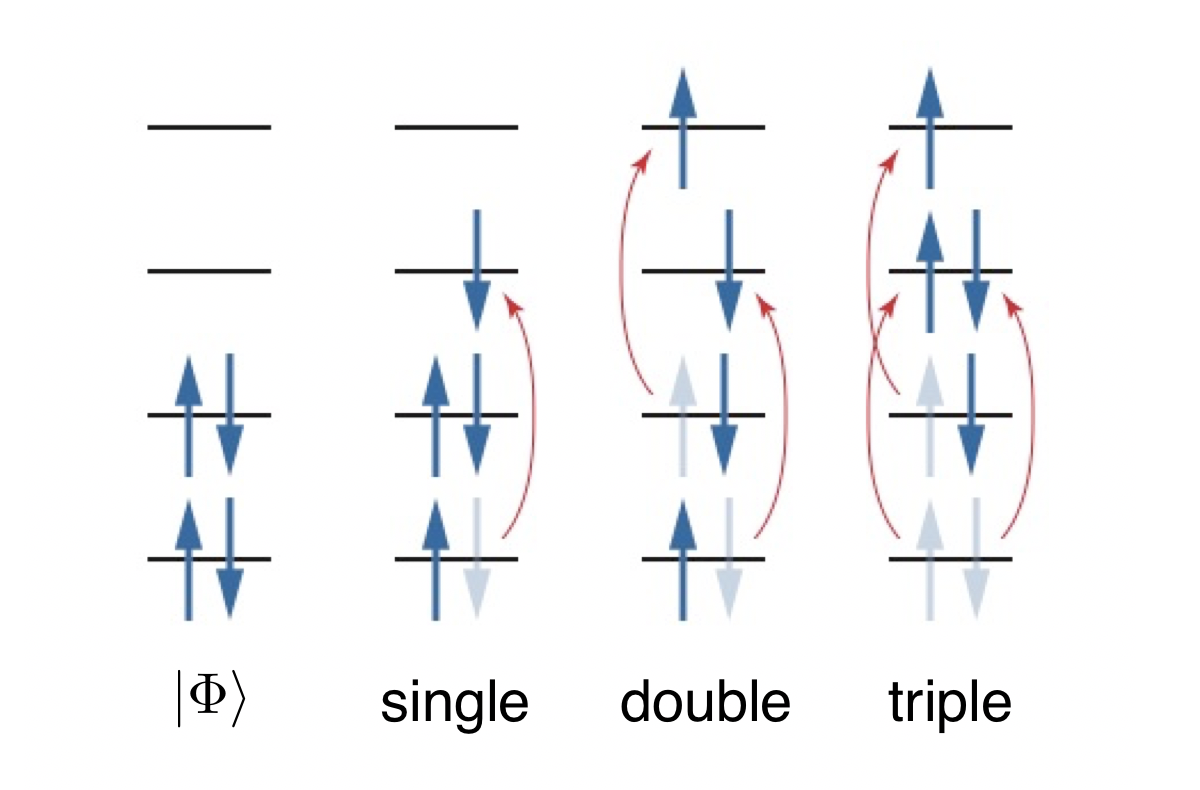}
    \caption{An example of a reference state $\ket{\Phi}$ and a single, double, and triple excitations from that state.}
    \label{fig:CCSD}
\end{figure}

Classical CC methods have been used to compute the impurity coupled cluster Green's function (CCGF)~\cite{nooijen92_55, nooijen93_15, nooijen95_1681} both for the AIM and other Green's function schemes for solving impurity models~\cite{PhysRevB.100.115154,JCTC.15.6010}. The CC approach solves the corresponding many-body problem by considering a subset of excitations from a reference state $\ket{\Phi}$~\cite{coester58_421, coester60_477, cizek66_4256, paldus72_50, purvis82_1910}. This method becomes exact in the limit that all possible excitations are considered; however, the computational complexity grows polynomially in the system size $N$ and factorially in the number of excited particles considered. Typical implementations use the Hartree-Fock solution as a reference and truncate the possible excitations to single-, double-, and sometimes triple-particle excitations (see. Fig.~\ref{fig:CCSD}). These implementations achieve $\mathcal{O}(N^6)$ for CCGF based on single and double excitations and $\mathcal{O}(N^8)$ when triple excitations are added. This scaling is a significant improvement over the exponential scaling of methods like exact diagonalizaton and its non-exact variants like the Lanczos algorithm, or quantum Monte Carlo methods applied to models with a sign problem. However, 
for strongly correlated regimes, where many-body effect become significant, one often needs to go beyond triple excitations to obtain converged solutions~\cite{Bartlett07RMP}. In this regime, classical algorithms for computing reliable CCGFs in either the frequency or time domain become severely constrained by memory and storage requirements~\cite{kowalski16_144101, kowalski18_4335, peng19_3185, vila20_6983}.

As with the classical CC algorithm, our hybrid quantum-classical algorithm 
for calculating the Green’s function requires a reference state $\ket{\Phi}$; however, we have found that we can obtain accurate results for the AIM using 
a simple product state. This aspect replaces the need for preparing a more complicated ground state on a quantum computer with a simpler task of applying unitary operators to a product state. To achieve this, we use the single CC amplitude obtained from fully converged classical CC calculations to construct the quantum measurement of the time-evolved unitary components of the non-Hermitian CCGF. This scheme allows us to extract the value of the Green’s function in the time domain. Using quantum computing algorithms for unitary time dynamics, our formalism can then calculate the impurity Green's function of our benchmark AIM model at time $t$ to precision $\varepsilon = \varepsilon_s + \varepsilon_m$
with a total $T$-gate cost of 
\begin{equation*}
\mathcal{O} \big([1-P_f]^{-1} \sqrt{\Upsilon} \varepsilon_s^{-1/2} \sqrt{t^3} N_\mathrm{bath}^5 \varepsilon_m^{-2}\big). 
\end{equation*}
Here, $P_f$ is the probability of failing to properly implement the linear combination of unitary operators, $\varepsilon_s$ is the synthesis error introduced by the time evolution operator, $t$ is the total evolution time, $N_{\text{bath}}$ is the number of continuum energy levels (or ``bath sites'') in the AIM, $\varepsilon_m$ is the error introduced by the measurement scheme used, and $\Upsilon = \frac{1}{12} \left[ \abs{ U_c} (\sum_i \abs{V_i})^2 + \frac{1}{2} U_c^2 \sum_i \abs{V_i} \right]$ is a constant dependent on the parameters of the AIM. 

\section{Methods} \label{sec:methods}

\subsection{The Anderson Impurity Model} \label{sec:AIM}
Our CC method is broadly applicable to many fermionic models of interest. Here, we demonstrate and benchmark our approach using the AIM~\cite{Anderson1961}, which has played an essential role in developing our understanding of localized fermionic degrees of freedom coupled to a continuum~\cite{AIMReview, Mahan1990}. The model describes localized magnetic moments embedded in a sea of conduction electrons and is closely related to the Kondo problem~\cite{Schrieffer1966, AIMReview}. It also plays a central role in dynamical mean-field theory (DMFT)~\cite{Georges1996}, where an infinite interacting system is mapped onto an AIM whose parameters are determined self-consistently to reproduce the physics of the original problem. (In the DMFT context, the continuum levels are often referred to as ``bath'' levels, and we will use this language throughout the paper.)

The Hamiltonian for the AIM is given by  
\begin{align} 
\begin{split}
    H = \sum_{i=0,\sigma}^{N_\mathrm{bath}} \epsilon^{\phantom\dagger}_i \hat{n}_{i,\sigma}^{\phantom\dagger} &+ U_c\hat{n}_{0,\uparrow}\hat{n}_{0,\downarrow} 
    \\& +  \sum_{i=1,\sigma}^{N_\mathrm{bath}} V^{\phantom\dagger}_i \left( c_{0 \sigma}^{\dagger} 
    c_{i, \sigma}^{\phantom\dagger} + \mathrm{H.c.} \right).\label{AM}
    \end{split}
\end{align}
Here, $i$ is an energy level index, where $i = 0$ corresponds to the impurity site and $i=1,\dots,N_\mathrm{bath}$ correspond to the energy levels of the continuum or bath, $U_c$ is the Coulomb repulsion on the impurity site, $V_i$ is the hybridization parameter that allows hopping between the bath and impurity levels, and $\epsilon_i$ are the on-site energies of each level. 

\subsection{Coupled Cluster Green's Functions from a sum of Unitary Operators}
Given a many-body system with a model Hamiltonian $H$, our goal is 
to construct the time-dependent single-particle retarded (advanced) Green's function $G_{pq}^{<}(t)$ [$G_{pq}^{>}(t)$] with the error of at most $\varepsilon > 0$. 
Here, we focus on the zero-temperature case, where the Green's functions 
are given by   
\begin{equation}
\text{i} G_{pq}^{<}(t) = \theta(t) 
\expval{\{ c^{\phantom{\dagger}}_{p}(t), 
c^{\dagger}_{q}(0) \}}{\text{GS}}
\end{equation}
and 
\begin{equation}
\text{i} G_{pq}^{>}(t) = \theta(t) 
\expval{\{ c^{\dagger}_{p}(t), 
c^{\phantom{\dagger}}_{q}(0) \}}{\text{GS}},
\end{equation}
where $c_{p}^{\phantom{\dagger}}(t)=e^{\mathrm{i}tH}c_{p}^{\phantom{\dagger}}(0)e^{-\mathrm{i}tH}$ is the fermionic annihilation operator in the Heisenberg representation and $|\rm{GS}\rangle$ is the ground state of $H$. 

Our algorithm to compute $G_{pq}^{<}(t)$ or $G_{pq}^{>}(t)$ consists of three main components, which address ground state preparation, time evolution, and measurements. 
As discussed in the introduction, the bottleneck for many quantum or quantum-
classical algorithms lies in preparing the ground state of the system. 
A novel aspect of our CC method is that it allows us to approximate the ground state by considering excitations from a simple product reference state $\ket{\Phi}$.  Specifically, the state $\ket{\rm GS}$ is approximated by applying cluster operators to $\ket{\Phi}$ (as in classical CC methods), but the operators are approximated here by a linear combination of unitary operators, as described in greater detail below.   


\subsection{State Preparation}
As described in Refs.~\citenum{nooijen92_55, nooijen93_15, nooijen95_1681}, the CCGF is built upon the CC  bi-variational exponential parametrization of the reference state $| \Phi \rangle$ for approaching the many-body ground-state wave function of a system 
\begin{equation}
\ket{\rm GS} = e^T \ket{ \Phi}
\end{equation}
and its dual 
\begin{equation}
\bra{\rm GS} = \bra{\Phi} (1+\Lambda) e^{-T},
\end{equation}
where $T$ and $\Lambda$ are cluster (excitation) and de-excitation operators, 
respectively. In the language of second quantization, $T$ and $\Lambda$ are given by sum of multi-particle scattering operators 
\begin{align}
\begin{split}
&T = \sum_{k=1}^{m} \frac{1}{(k!)^2} \sum_{\substack{i_1,\ldots,i_k ; \\ a_1, \ldots, a_k}} t^{i_1\ldots i_k}_{a_1\ldots a_k} c_{a_1}^\dagger\cdots c_{a_k}^\dagger c^{\phantom\dagger}_{i_k} \cdots c^{\phantom\dagger}_{i_1} , \\
& \Lambda = \sum_{k=1}^{m} \frac{1}{(k!)^2} \sum_{\substack{i_1,\ldots,i_k ; \\ a_1, \ldots, a_k}} \lambda_{i_1\ldots i_k}^{a_1\ldots a_k} c_{i_1}^\dagger\cdots c_{i_k}^\dagger c^{\phantom\dagger}_{a_k} \cdots c^{\phantom\dagger}_{a_1} . \label{L}
\end{split}
\end{align}
Here, the indices $i_1,i_2,\ldots$ ($a_1,a_2,\ldots$) denote occupied (unoccupied) spin-orbitals in the reference $|\Phi\rangle$, the coefficients $t^{i_1\ldots i_k}_{a_1\ldots a_k}$'s and $\lambda_{i_1\ldots i_k}^{a_1\ldots a_k}$'s are scalar amplitudes, and $m$ is the excitation level ($\le$ number of electrons) that defines the approximation in the CC hierarchy (e.g. $m=2$ corresponds to CC singles and doubles (CCSD)~\cite{purvis82_1910}, $m=3$ corresponds to CC singles, doubles, and triples (CCSDT)~\cite{ccsdt_noga,ccsdt_noga_err,SCUSERIA1988382}, etc., see Fig.~\ref{fig:CCSD}). 

\begin{figure*}[t]
    \centering
    \includegraphics[width=0.7\linewidth, keepaspectratio]{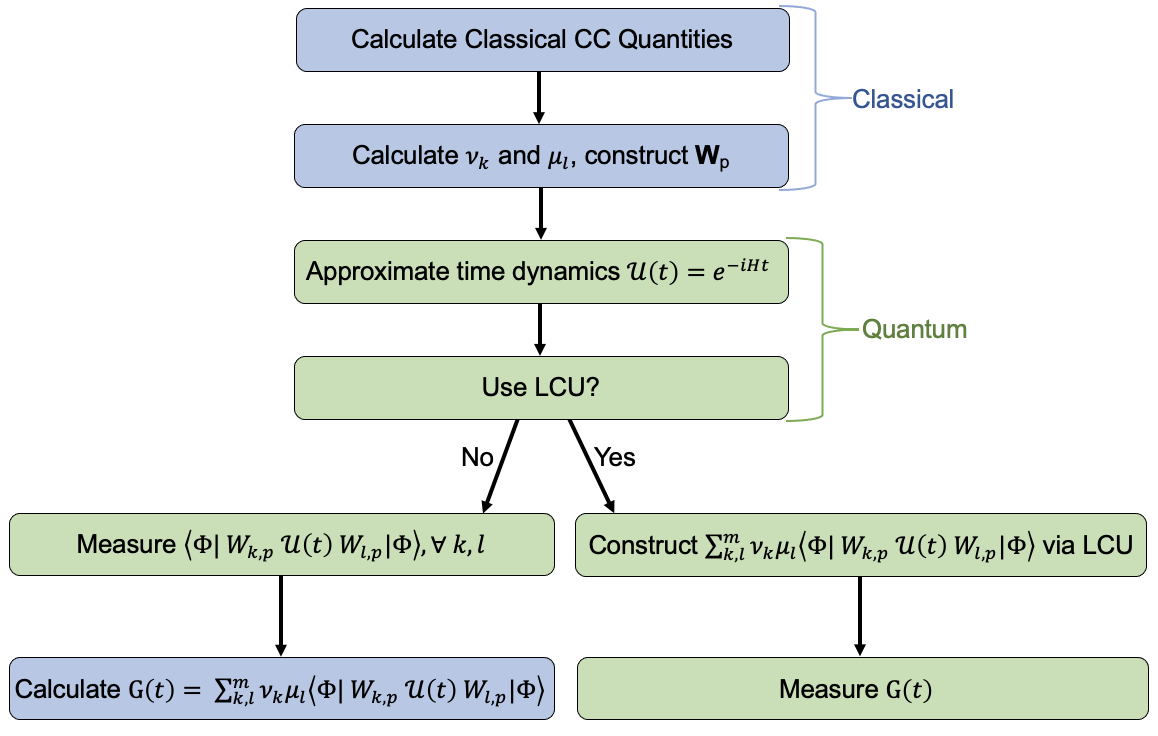}
    \caption{An overview of the quantum-classical coupled-cluster algorithm for computing the Green's function at local site $p$ of a fermionic model. Boxes in blue (green) indicate tasks completed on classical (quantum) computer hardware. $\mathbf{W}_p$ is the set of unitaries obtained using the CCGF fermion-to-unitary mapping.}
    \label{fig:algorithm}
\end{figure*}

By employing the CC bi-variational parametrization, the time-dependent CCGF for a many-body electron system can be expressed as
\begin{equation} 
G_{pq}(t) =G^{<}_{pq}(t) + G^{>}_{pq}(t),\label{gfxn_t}
\end{equation}
where 
\begin{equation} \label{gfxn<}
G^{<}_{pq}(t)=\langle (1+\Lambda) e^{-T} c_q^{\dagger} e^{-\text{i}2\pi(H-E_{\rm CC})t} c_p e^T \rangle
\end{equation}
and
\begin{equation} \label{gfxn>}
G^{>}_{pq}(t)=\langle (1+\Lambda) e^{-T} c_p e^{+\text{i}2\pi(H-E_{\rm CC})t} c_q^{\dagger} e^T \rangle.
\end{equation}
Here, $\langle O \rangle=\langle \Phi | O | \Phi \rangle$ denotes the expectation value of an operator $O$ with respect to the reference $|\Phi\rangle$. In the classical CCGF formulation, the operators appearing in Eqs.~\eqref{gfxn<} and ~\eqref{gfxn>} are not unitary. This method must be modified for use on quantum computers, which require unitary operations.

We now illustrate how to develop a unitary formulation of the CC method using the lesser  part $G^{<}_{pq}(t)$ [Eq. ~\eqref{gfxn<}]. Similar considerations apply to the greater term $G^{>}_{pq}(t)$. We have found that at least one time-independent set of $m$ unitaries $\mathbf{W}_p = \{ W_{1,p}, W_{2,p}, \cdots, W_{m,p} \}$ can be employed to expand 
$c_p e^{T} | \Phi \rangle = \sum_i \mu_i W_{i,p} | \Phi \rangle$ and
$\langle \Phi |(1+\Lambda) e^{-T} c_q^\dagger = \sum_i \nu_i \langle \Phi | W_{i,q}^\dagger$ 
with the scalar coefficients $\{\mu_i\}$ and $\{\nu_i\}$ such that Eq. (\ref{gfxn<}) can be fully unitarized. If we limit our discussion to the CCSD framework (i.e. $T \approx T_1+T_2 = \sum_{ai}t_i^a \hat{E}^a_i + \sum_{a<b,i<j} t_{ij}^{ab} \hat{E}^{ab}_{ij}$, where $\hat{E}^{ab\cdots}_{ij\cdots} = c_a^\dagger c_b^\dagger \cdots c^{\phantom\dagger}_j c^{\phantom\dagger}_i$ for $a,b\in S$ and $i,j\in O$, with $S$ and $O$ being 
the virtual and occupied subspace, respectively) and the index $p\in O$, then one option for $\textbf{W}_p$ is (see detailed derivation in \ref{sec:ccstuff})
\begin{eqnarray*}
\mathbf{W}_p&=&\big\{ 
\{\tilde{X}_p\}, 
\{ \tilde{X}_a \tilde{X}_i \tilde{X}_p\}_{\substack{a; \\ i\neq p}}, \\\nonumber 
&&\quad\quad\quad
\{\tilde{X}_a \tilde{X}_b \tilde{X}_j \tilde{X}_i \tilde{X}_p\}
_{\substack{a<b,i<j; \\ i\neq p, j\neq p}} 
\big\}. 
\end{eqnarray*}
Here, $\tilde{X}_i$ denotes an $X$ Pauli operator (or gate) acting on the $i$\textsuperscript{th} spin-orbital (or qubit) with $Z$ Pauli operators acting on all qubits before it in our ordering, and $m \sim \mathcal{O}(N_S^2N_O^2) \sim \mathcal{O}(N_{\rm bath}^4)$, where $N_S$ and $N_O$ denote the numbers of virtual and occupied spin-orbitals, respectively.
The coefficients $\{\mu_i\}$ and $\{\nu_i\}$ are given by cluster amplitudes (see \ref{sec:ccstuff}). Note that the CCSD ground state energy for a general AIM with one impurity site only depends on the single amplitude, i.e. $\Delta E_{\text{CCSD}}=\sum_{i=1}^{N_{\text{bath}}}V^{\phantom\dagger}_i t_0^i$ since the double amplitudes in a CCSD calculation of AIM are only used to converge the single amplitudes corresponding to the excitation from the impurity to the bath. Thus, if we only use the converged $T_1$ amplitude to construct (or approximate) the correlated wave function, the associated ground state energy will be exactly same as the CCSD correlation energy but the number of unitary vectors $m$ is greatly reduced from $\mathcal{O}(N_{\text{bath}}^4)$ to $\mathcal{O}(N_{\text{bath}})$. 

Once the unitary set $\mathbf{W}_p$ and time propagator operator $\mathcal{U}(t)$ are constructed, the lesser CCGF can be expressed as
\begin{equation} 
G^{<}_{pq}(t) = \sum_{k,l}^m \nu_k \mu_l
\langle W^\dagger_{k,q} \mathcal{U}(t) W^{\phantom\dagger}_{l,p} \rangle, \label{lesser}
\end{equation}
where the expectation values are computed using a quantum device. In the case of the AIM, the lesser CCGF for the impurity site would correspond to $p = q = 0$. 
Note that the gate depth for $W_{i,p}$ is just $\mathcal{O}(1)$, while the number of terms in Eq.~(\ref{lesser}) scales as $\mathcal{O}(m^2)$. 
 The method(s) for constructing the time evolution operator $\mathcal{U}(t)$ and measuring the expected values in Eq.~(\ref{lesser}) are discussed in the next two subsections.

In analogy to the single-reference CC formulations discussed above, one can map to $\mathbf{W}_p$ other parametrizations of the ground wave function. 
For example, one can consider a unitary representation 
\begin{equation}
c_p e^{\tau}|\Phi\rangle \;, \label{ucc}
\end{equation}
where $\tau$ is an anti-Hermitian cluster operator, or multi-reference CC expansions 
\begin{equation}
c_p \sum_{\mu=1}^M \alpha_{\mu} e^{T^{(\mu)}} |\Phi_{\mu}\rangle \;,
\label{mrcc}
\end{equation}
where the coefficients $\alpha_{\mu}$ define the eigenvector of the effective Hamiltonian, 
$\lbrace |\Phi_{\mu}\rangle \rbrace$ are reference functions defining model space, and the operators $T^{(\mu)}$ are reference-specific cluster operators. If we choose a specific Slater determinant $|\Phi_{\nu}\rangle$ as a reference, other Slater determinants 
$|\Phi_{\mu}\rangle$ can be obtained as a $|\Phi_{\mu}\rangle=\Omega_{\mu\nu} |\Phi_{\nu}\rangle \;\; \forall_{\mu \ne \nu}$, where the operator $\Omega_{\mu\nu}$ contains a string of creation/annihilation operators carrying active spin-orbital indices only.

\subsection{Time evolution}
There are multiple methods for approximating the time evolution operator. One  
option, which we adopt here, is to use a symmetrized Trotter formula with a Givens rotation, as described in Ref.~\citenum{Kivlichan2020}. In this case, we separate the 
potential $(U_c \text{ and } \epsilon)$ terms from the hopping $(V)$ terms and implement them separately as $\mathcal{U}(t)\approx e^{-\text{i} \frac{t}{2} H_{\text{pot}}} e^{-\text{i} t H_{\text{hop}}} e^{-\text{i} \frac{t}{2} H_{\text{pot}}}$, as proposed in Ref.~\citenum{Kivlichan2020}. By implementing a series of Givens rotations between the potential and hopping terms, the hopping terms become single qubit rotations, which must be synthesized using $T$ gates.
Note that this Trotter decomposition will require two implementations of the set of Givens rotations, a set of rotations into the diagonal basis of the hopping Hamiltonian and a set of rotations back into the computational basis.

For more advanced simulation methods such as qubitization\cite{Low2019} and Taylor series expansion~\cite{BerryTaylor2015}, oracles are required. They incur a cost based on query complexity, i.e. the number of times an oracle must be accessed to perform the simulation. These oracles must also be constructed using basic gates and ancilla qubits, thus increasing the gate count. However, it was recently shown that the most expensive oracle to implement in the Taylor series and Qubitization methods, the Select(H) oracle, can be implemented with just $\mathcal{O}(N)$  gates\cite{Wan2021}. For a comparison of time evolution methods in terms of gate count, see Tab. \ref{tab:scaling}.

\setlength\tabcolsep{0pt}
\renewcommand{\arraystretch}{2.1}

\newcommand{\specialcell}[2][c]{%
\begin{tabular}[#1]{@{}c@{}}#2\end{tabular}}

\begin{table*}[t]
\centering
\resizebox{\textwidth}{!}{%
\begin{tabular}{cccccc}
\hline
\rowcolor[HTML]{EFEFEF} 
\textbf{Algorithm} & \textbf{Ref.} & \textbf{Ancilla Qubits}& 
\textbf{Query 
Complexity}  & ~\textbf{Gate/Query} & \textbf{Total Gate Count } \\ \hline
~Trotter (Givens rotation)~    &~\citenum{Kivlichan2020} & 0 & - & - &$\quad\mathcal{O} \bigg( \sqrt{\Upsilon} \epsilon_s^{-1/2} \sqrt{t^3} N_\mathrm{bath} \log N_\mathrm{bath} \bigg)\quad$     
 \\ \hline
\rowcolor[HTML]{EFEFEF} 
Taylor Series          &~\citenum{BerryTaylor2015,Wan2021} & $\quad\mathcal{O} 
\big( \log N_\mathrm{bath}  f(\norm{\vec{\alpha}}_1 \tfrac{t}{\epsilon_s})\big)\quad$ & $\mathcal{O}\big( \norm{\vec{\alpha}}_1 t f(\norm{\vec{\alpha}}_1 \tfrac{t}{\epsilon_s})\big) $ & $\quad\mathcal{O}\big(N_\mathrm{bath} \big)\quad$  & $\mathcal{O} 
\big( \norm{\vec{\alpha}}_1 
t N_\mathrm{bath}  f(\norm{\vec{\alpha}}_1 \tfrac{t}{\epsilon_s})\big)$   \\ \hline
Qubitization &~\citenum{Low2019, Wan2021} & $\ceil*{\log{N_\mathrm{bath}}}+2$ & $\mathcal{O} \big(\norm{\vec{\alpha}}_1 t + f(\tfrac{1}{\epsilon_s}) \big)$ & $\mathcal{O} (N_\mathrm{bath})$  & $\mathcal{O}\big( N_\mathrm{bath} (\norm{\vec{\alpha}}_1 t + f(\tfrac{1}{\epsilon_s})) \big)$\\
\hline
\end{tabular}%
}
\caption{Number of required ancillae and T-gate count scaling as a function of allowed error in time evolution $\epsilon_s$, total evolution time $t$, and the number of bath sites $N_{\text{bath}}$ for three common time evolution procedures. Here, we take $H=\sum_{i=1}^L \alpha_i H_i$, $f(\eta) = \tfrac{\log \eta }{\log\left( \log \eta\right) }$, $\Upsilon = \frac{1}{12} \left( \abs{ U_c} (\sum_i \abs{V_i})^2 + \frac{1}{2} U_c^2 \sum_i \abs{V_i} \right)$.}
\label{tab:scaling}
\end{table*}

\subsection{Measurement}
Although the scaling of the number of measurements needed to calculate the Green's function grows \emph{quadratically} in the system size, the linear combination of the expectation values needed to calculate the impurity Green's function can be calculated in one circuit using the linear combination of unitaries (LCU) method\cite{Childs2012}. 

To implement a linear combination of $k+1$ unitary operators, LCU requires $\log(k)$ ancillary qubits. Using the LCU method to measure the Green's function of the AIM thus requires  $k\sim\mathcal{O}(N_{\text{bath}}^2)$, and  the number of required ancilla qubits therefore grows as  $\log(N_{\text{bath}}^2)$. From theorem 3 of Ref.~\citenum{Childs2012}, we know the probability of success for this method. Consider a linear combination of unitary
operators $R = \sum\limits_{q=1}^{k+1} C_q U_q$, with $k\geq1$, $\norm{U_q} = 1$, and $\max_{q \neq q^\prime} 
\norm{U_q - U_{q^\prime}} \leq \Delta$. Let $\kappa = \sum_{q:C_q>0}C_q / \sum_{q:C_q<0} \abs{C_q}$. Then there exists a quantum algorithm that implements an operator proportional to $R$ with failure probability $P_f = P_+ + P_-$, where $P_+ \leq {\kappa \Delta^2}/{4}$ and $P_- \leq {4\kappa}/{(\kappa+1)^2}$ are the probabilities of failure to add and subtract two terms, respectively.
We utilize this method to measure each term in Eq.~(\ref{lesser}) with one circuit, thus reducing the measurement error.

To measure the entire Green's function in the time-domain with just one circuit with LCU, we first construct the circuit that implements the LCU that is comprised of all of the terms in Eq.~(\ref{lesser}), and then measure the expectation value of the LCU operator with the Hadamard test. It is also possible to measure each expectation value in Eq.~(\ref{lesser}) by using the Hadamard test with each term in an independent circuit, and then combine them. The $T$-gate counts for the LCU method and the direct Hadamard test are given in the Error Analysis and Gate Count Scaling section. Because the LCU method has a failure probability, it is possible that the direct Hadamard test could be advantageous in terms of gate count, even though it has poorer scaling with respect to the system size.

\begin{widetext}
We now focus on computing the local Green's function $G_{pp}(t)$ of the AIM. 
In this case, the circuit to measure the real part of $G_{pp}(t)$ in the time-domain via LCU is
\begin{equation*}
\hspace{-5cm}
           \Qcircuit @C= 1.0em @R=1.3em {
            \lstick{\ket{+}}    & \qw & \ctrl{1} & \qw & \gate{H} & \meter & \rstick{\mathrm{Re}[\bra{\Phi}\sum_{k,l}^m \nu_k \mu_l W_{k,p} \mathcal{U}(t) W_{l,p} \ket{\Phi} ]} \\
            \lstick{\ket{0}} & \qw  {/}  & \multigate{1}{\sum_{k,l}^m \nu_k \mu_l W_{k,p} \mathcal{U}(t) W_{l,p}} & \qw & \qw & \qw   \\
            \lstick{\ket{\Phi}}& \qw  {/}   & \ghost{\sum_{k,l}^m \nu_k \mu_l W_{k,p} \mathcal{U}(t) W_{l,p}}& \qw & \qw & \qw 
}\end{equation*}
\end{widetext}

where the $\order{\log(N_{\text{bath}}^2) }$ qubit state of the ancilla register $\ket{0}$ is prepared in a state determined by the coefficients $\nu_k \text{ and } \mu_l$ of the LCU. The top single ancilla qubit is the one utilized for the Hadamard test. Finally, to measure the imaginary part of the expectation value instead of the real part as shown, one needs only to apply two single qubit gates to the ancilla qubit.

\begin{figure*}[t!]
    \centering
    \includegraphics[width=0.75\textwidth]{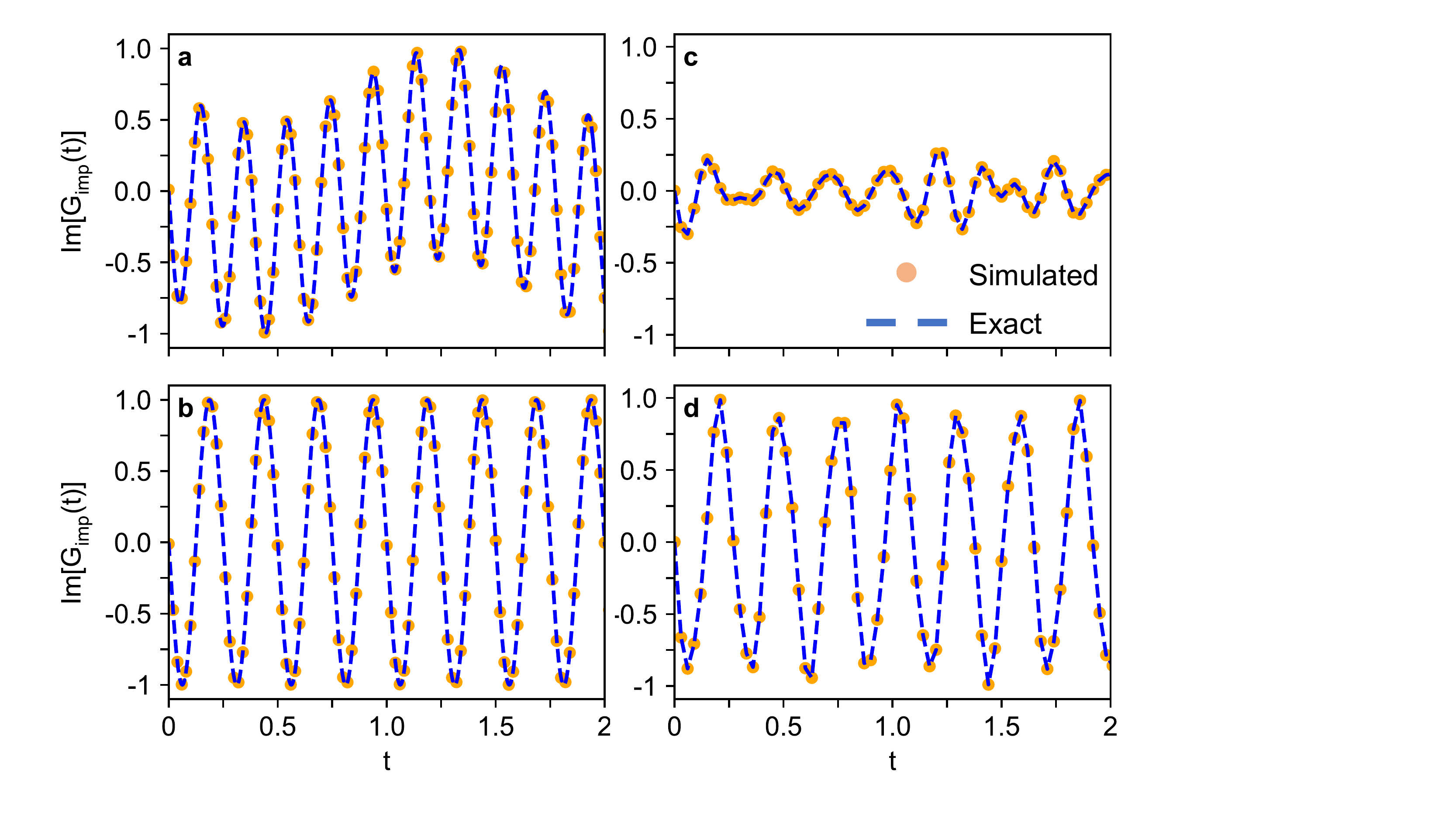}
    \caption{Impurity Green's function in the time domain for the AIM model with $U=8$. The left panels show results for \textbf{a}~$N_\mathrm{bath} = 1$, $\epsilon_i = \{4,0\}$, and $V_1=1$, and \textbf{b}~$N_\mathrm{bath} = 1$, $\epsilon_i = \{4, 0\}$, and $V_1=0$. The right panels show results for 
    \textbf{c}~$N_\mathrm{bath} = 2$, $\epsilon_i = \{4, 3.61, 4.39\}$, and $V_i = \{0.63, 0.63\}$, and \textbf{d}~$N_\mathrm{bath} = 2$, $\epsilon_i = \{4,-0.13, 10.1\}$, and $V_i = \{1, 0.15\}$, 
    These parameter values were selected to be representative of values that would typically arise when solving the AIM in the context of a dynamical mean-field theory algorithm. 
    }
    \label{fig:gimpt}
\end{figure*}

\section{Results and Benchmarks} \label{sec:results}


\subsection{Two and Three-Site Examples}
To demonstrate our method, we employed this hybrid quantum-classical algorithm to compute the many-body CCGF of AIM with $N_\mathrm{bath} = 1$ and $2$ and compared our results with the exact solutions obtained with exact diagionalization. 

For the $N_\mathrm{bath} = 1$ case, we employ four qubits on the quantum simulator within the Qiskit framework\cite{Qiskit}. We represent the model system with qubit \#1 and \#3 denoting the impurity site, and use a simple reference state $|\Phi\rangle = |0110\rangle$. For the lesser CCGF, the set $\mathbf{W}_p$ ($p$ = 3) then only includes two elements $\{ \tilde{X}_3$, $\tilde{X}_1\tilde{X}_2\tilde{X}_3  \}$, such that $G^{<}_{\text{imp}}$ is given by a linear combination of only three terms
\begin{eqnarray}
G^{<}_{\text{imp}} &\Leftarrow & \Big\{\langle \tilde{X}_3\mathcal{U}(t)\tilde{X}_3\rangle, \langle \tilde{X}_3\mathcal{U}(t)\tilde{X}_1\tilde{X}_2\tilde{X}_3\rangle, \notag \\
& &\langle \tilde{X}_3\tilde{X}_2\tilde{X}_1\mathcal{U}(t)\tilde{X}_1\tilde{X}_2\tilde{X}_3\rangle \Big\}
\end{eqnarray}
with the coefficient for each term determined from the product of the elements of $\{\mu_i\}$ and $\{\nu_i\}$ ($i=1,2$) as demonstrated in Eq. (\ref{lesser}). The same strategy can be easily extended to an AIM with more bath levels. For example, for simulating a three-level model, six qubits can be employed to represent the model system with two qubits denoting the impurity site, and the trial state given by $|110010\rangle$. In this case, computing the lesser CCGF Green's function requires a set $\mathbf{W}_p$ ($p$ = 3) with six elements $\{ \tilde{X}_3$, $\tilde{X}_5\tilde{X}_1\tilde{X}_3$, $\tilde{X}_4\tilde{X}_2\tilde{X}_3$, $\tilde{X}_6\tilde{X}_2\tilde{X}_3$, $\tilde{X}_5\tilde{X}_4\tilde{X}_2\tilde{X}_1\tilde{X}_3$, $\tilde{X}_5\tilde{X}_6\tilde{X}_2\tilde{X}_1\tilde{X}_3  \}$, such that  $G^{<}_{\text{imp}}$ is given by a sum over twenty-one terms. However, based on our previous finding, these twenty-one terms can be reduced to just three terms related to only two elements of the set $\mathbf{W}_p$,  $\{\tilde{X}_3,\tilde{X}_4\tilde{X}_2\tilde{X}_3\}$. This reduction is due to the fact that the CCSD ground state energy for the AIM only depends on the single excitation cluster amplitudes.

To implement the time evolution operator in our examples, we separate the 
potential $(U_c \text{ and } \epsilon_i )$ terms from the hopping $(V_i)$ terms and implement the former as a second order symmetric product formula. However, 
for our two and three-site examples,
we do not implement the more advanced Trotter method with Givens rotations and 
instead implement the terms directly. 

\begin{figure*}[t!]
    \centering
    \includegraphics[width=0.7\textwidth]{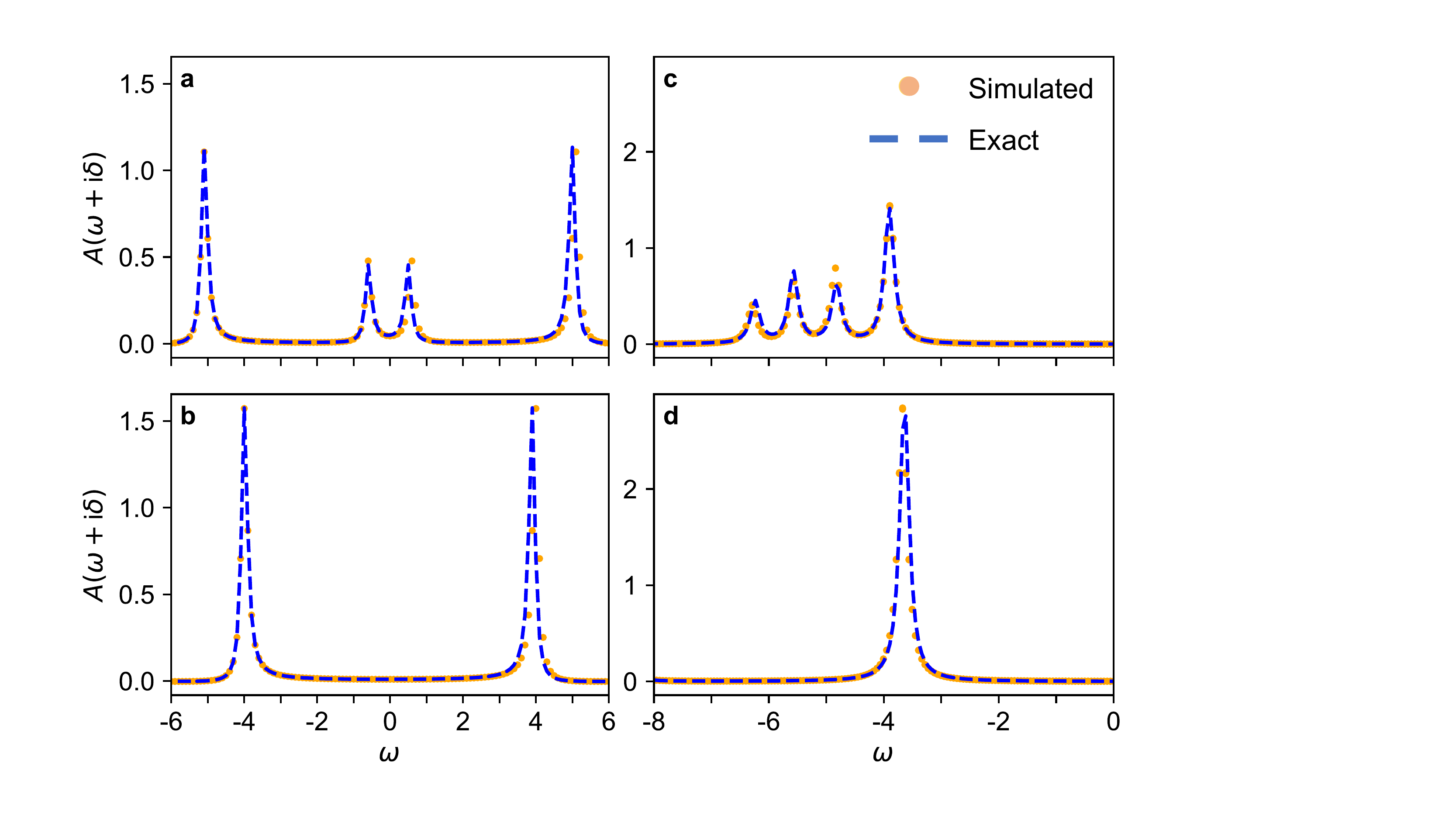}
    \caption{The local spectral function of the impurity site of the AIM model with $U=8$. The left panels show results for \textbf{a}~$N_\mathrm{bath} = 1$, $\epsilon_i = \{4,0\}$, and $V_1=1$, and \textbf{b}~$N_\mathrm{bath} = 1$, $\epsilon_i = \{4, 0\}$, and $V_1=0$. The right panels show results for 
    \textbf{c}~$N_\mathrm{bath} = 2$, $\epsilon_i = \{4,3.61, 4.39\}$, and $V_i = \{0.63, 0.63\}$, and \textbf{d}~$N_\mathrm{bath} = 2$, $\epsilon_i = \{4,-0.13, 10.1\}$, and $V_i = \{1, 0.15\}$, 
    These parameter values were selected to be representative of values that would typically arise when solving the AIM in the context of a dynamical mean-field theory algorithm. 
    }
    \label{fig:spec}
\end{figure*}

Figure \ref{fig:gimpt} shows the impurity Green's function in the time 
domain for the two- (Fig.~\ref{fig:gimpt}{\bf a} and \ref{fig:gimpt}{\bf b}) and three-site (Fig.~\ref{fig:gimpt}{\bf c} and \ref{fig:gimpt}{\bf d}) AIM. 
(In both cases, our simulations were conducted for parameters that typically arise in the DMFT self-consistency loop, along with a parameter set that corresponds to the final self-consistent solution.) We then fast-fourier transform (FFT) the time domain impurity Green's function to the frequency domain to obtain the spectral functions $A_\mathrm{imp}(\omega)=-\mathrm{Im}G_\mathrm{imp}(\omega + \mathrm{i}\delta)/\pi$, as shown in Fig. \ref{fig:spec}. For the three-site system, we tested the dependence of the solution on the number of Trotter steps taken per time step. We found that the results shown in Figs. \ref{fig:gimpt}{\bf c} and \ref{fig:gimpt}{\bf d} had little dependence on the number of Trotter steps taken for our choice of parameters and time step size. For example, data obtained for 8 and 32 Trotter steps per time step essentially coincide in Figs \ref{fig:gimpt}{\bf c} and \ref{fig:gimpt}{\bf d}. Here, the insensitivity of the solution on the number of Trotter steps is due to the small time step size we employed. In all cases, our hybrid quantum-classical algorithm reproduces the exact solution obtained by exact diagonalization. 

Although the examples of the two- and three-site Anderson impurity model are relatively simple, it is important to note that this method is very general. Here, we focused on the solutions of the two- and three-site problems for two sets of parameters (each) in order to emphasize that the efficacy of our approach does not depend strongly on the specific model parameters, even though the specifics of the model lead to simplifications in the method. It is also important to see the flexibility of the $\mathbf{W}_p$ operators in the construction of any form of the $N$-electron ground state wave functions.

\section{Discussion} \label{sec:discussion}
\subsection{Advantages over VQE-based methods}
Several VQE-based methods to calculate the Green's function have been proposed in the literature. Ref.~\citenum{Kosugi2020}, for example, uses a unitary coupled cluster ansatz to prepare an approximation to the ground state, which is then used to generate transition matrix elements via statistical sampling. These transition matrix elements are then used in conjunction with the (known) pole locations (obtained via any one of a number of previously proposed algorithms for obtaining many-electron energy eigenvalues) to construct the Green's function of the lithium hydride and water molecules. In a similar variational formulation, the authors of Ref.~\citenum{Endo2020} utilize the VQE framework to calculate both the real-time Green's functions and the Lehmann representation of the spectral function of a many-body system. To do this, the authors utilize VQE to construct the ground state, implement time evolution, find transition matrix elements, and obtain the many-electron energy eigenvalues needed in the Lehmann representation. As a numerical demonstration, they utilize the two-site Fermi-Hubbard model.

Our method differs from these VQE-based methods in many important ways. First, we do not require any prior determination of the many-electron energy eigenvalues. Instead, we work directly in the time domain and use a Fourier transform to obtain them. Second, our method does not rely on classical optimization algorithms, thus bypassing the so called ``barren minima'' problem that can occur in VQE solutions. Third, we have flexibility in the definition of the time-independent ${\bf W}_p$ sets, which offers a tuning mechanism to the available quantum resources. The ${\bf W}_p$ sets also allow for the definition of selective sub-sets of excitations to describe correlations in the $(N \pm 1)$-electron space. This freedom could be exploited in the construction of a subspace representation of the Green's function, as is done in active-space formulations of ionization-potentials/electron-affinities equation-of-motion coupled-cluster methods.

Finally, the ${\bf W}_p$ sets also provide great flexibility in emulating any form of the $N$-electron ground-state wave functions, including higher-order coupled-cluster wave function expansions, configuration interaction representations, and multi-reference wave functions that may be required to handle strong correlation effects in the ground state. In this paper, we focused only on the single-reference CC parametrizations of the ground-state wave functions. From this perspective, the possibility of mapping arbitrary wave functions into the Pauli strings spanning the ${\bf W}_p$ subspace defines the universal character of the proposed quantum algorithm in dealing with various many-body systems.

\subsection{Error Analysis and Gate Count Scaling} \label{sec:gatecount}
Suppose we want to calculate the impurity Green's 
function of an AIM with $N_{\text{bath}}$ bath sites in the time domain, to 
within a total error
$\varepsilon$ for a total time $t$. We consider two main 
contributions to 
$\varepsilon$, namely the synthesis error $\varepsilon_s$
from implementing the time evolution operator, and 
$\varepsilon_m$ due to statistical errors in the 
measurement scheme. Here, we propose utilizing 
LCU with the Hadamard test to measure the entire 
impurity Green's function 
in the time domain with one circuit, using the Trotter (Givens rotations) 
method to approximate the time evolution operator. 
Alternatively, one could 
measure each expectation value in Eq.~\eqref{lesser}
separately with the Hadamard test, again using
the Trotter (Givens rotation) method of Ref.~
\citenum{Kivlichan2020} to approximate the time 
evolution operator. In our two-site example, we use
the Hadamard test method for the convenience of implementation.

With the LCU method, for a given failure probability $P_f$, measurement error tolerance $\varepsilon_m$, and number of gates $N_g$ required to implement the multiple controlled unitary operators, we need to use $(1-P_f)^{-1} 
N_g \varepsilon_m^{-2}$ gates to successfully measure the full impurity 
Green's function. Here, we use the convention that the variance in a 
single measurement is $\varepsilon_m$, and grows as $\varepsilon_m = 1/\sqrt{N_s}$ with $N_s$ the number of samples taken. In this case, the number
of ancilla qubits required to implement the LCU will grow as $\mathcal{O}
(\log N_{\text{bath}}^2 )$, plus one ancilla for the Hadamard test. Breaking the multiple controlled
qubits into basis gates requires extra "work qubits," and the number of work qubits needed grows as the number of ancillas in the control register minus one. This additional requirement does not change the asymptotic scaling of the number of ancilla qubits required; it remains
$\mathcal{O}(\log N_{\text{bath}}^2 )$.
(See, e.g. Ref.~\citenum{NielsenChuang} for gate construction of the multiple controlled unitaries.) 

\begin{figure*}[t!]
    \centering
    \includegraphics[width=0.75\textwidth, keepaspectratio]{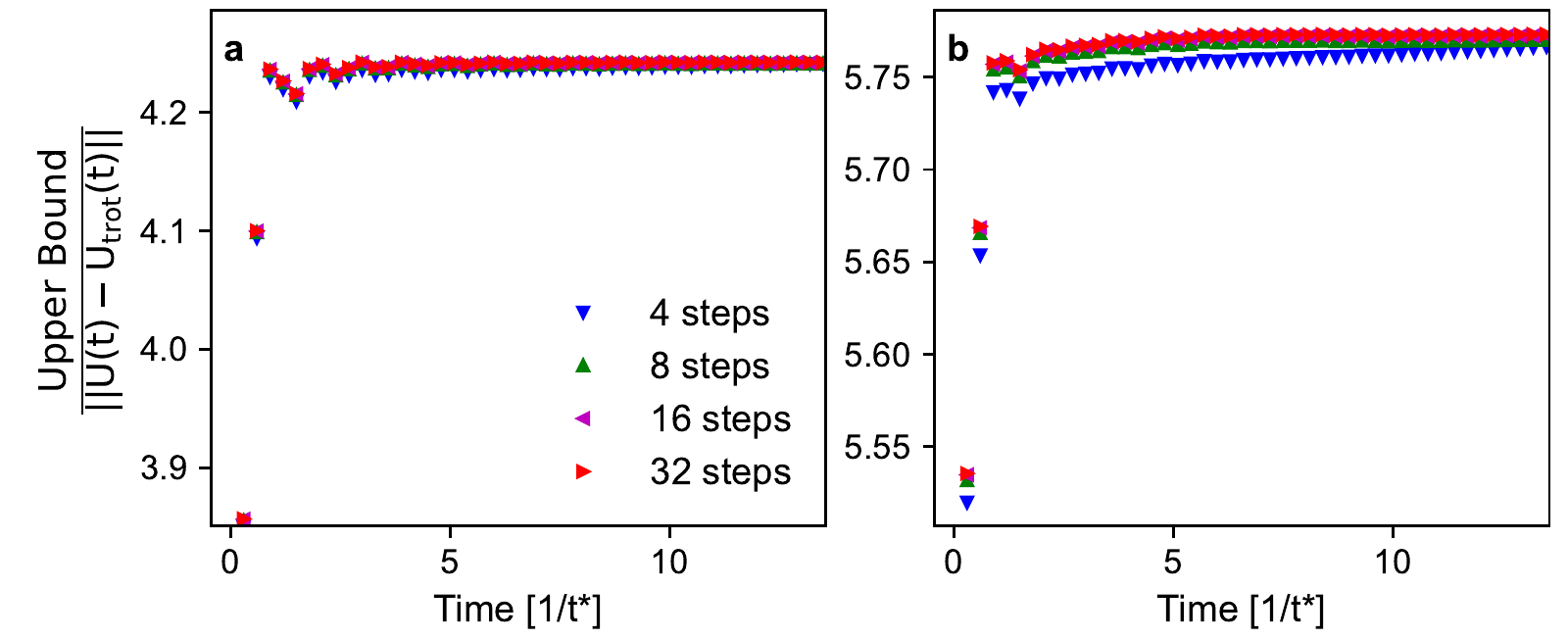}
    \caption{Ratio of the upper bound on the Trotter error from Ref.~\citenum{Kivlichan2020} to the actual Trotter error for \textbf{a} ~$N_\mathrm{bath} = 1$, $\epsilon_i = \{4,0\}$, and $V_1=1$ and \textbf{b}~$N_\mathrm{bath} = 2$, $\epsilon_i = \{4, 3.61, 4.39\}$, and $V_i = \{0.63, 0.63\}$. For testing purposes, the size of the time steps was set to $0.03$ and the number of Trotter steps used per time step is given in the legend.}
    \label{fig:trot}
\end{figure*}

To implement the multiple controlled version of each unitary within the LCU method, $\mathcal{O}(\sqrt{\Upsilon}\varepsilon_s^{-1/2} \sqrt{t^3} N_{\text{bath}}^2)$ gates are 
required, where $t$ is the total 
evolution time and $\Upsilon = \frac{1}{12} \left( \abs{ U_c} (\sum_i \abs{V_i})^2 + \frac{1}{2} U_c^2 \sum_i \abs{V_i} \right)$ is a system-dependent factor from utilizing the Trotter decomposition of the time evolution operator. Implementing all $\mathcal{O} (N_{\text{bath}}^2)$ multiple 
controlled unitaries then requires $\mathcal{O}(\sqrt{\Upsilon}\varepsilon_s^{-1/2} 
\sqrt{t^3} N_{\text{bath}}^4)$ gates. Note, however, that  
\emph{each} unitary will have some synthesis error due to the approximation of 
the time evolution operator and there will be $\mathcal{O}(N_{\text{bath}}^2)$
time evolution operators in the full circuit. Therefore, to maintain a maximum synthesis error of $\varepsilon_s$, each time evolution operator can contribute no more than $\varepsilon_s/N_{\text{bath}}^2$ error. In the above expression, substituting 
$\varepsilon_s \to \varepsilon_s/N_{\text{bath}}^2$ gives a total gate scaling of 
$\mathcal{O}(\sqrt{\Upsilon}\varepsilon_s^{-1/2} \sqrt{t^3} N_{\text{bath}}^5)$. Thus, we have a full gate count of $\mathcal{O}([1-P_f]^{-1} \sqrt{\Upsilon} \varepsilon_s^{-1/2} 
\sqrt{t^3} N_{\text{bath}}^5 \varepsilon_m^{-2})$ to measure the impurity 
Green's function in the time domain. This implementation has
the advantage of only requiring one circuit to measure
with the Hadamard test.

If one uses the Hadamard test to measure each term in Eq.~(\ref{lesser}) separately, each controlled unitary will require $\mathcal{O}(\sqrt{\Upsilon} \varepsilon_s^{-1/2} \sqrt{t^3} N_{\text{bath}}^2)$ gates. Since there are $\mathcal{O}(N_{\text{bath}}^2)$ controlled unitaries to be implemented, this yields a gate complexity of $\mathcal{O}(\sqrt{\Upsilon} \varepsilon_s^{-1/2} \sqrt{t^3} N_{\text{bath}}^4)$. Since each unitary includes time evolution with some synthesis error, we again take $\varepsilon_s \to \varepsilon_s/N_{\text{bath}}^2$. Similarly, for each of the unitary operators to be measured, the measurement error $\varepsilon_m$ grows as $1/\sqrt{N_s}$, where $N_s$ is again the number of samples taken. Thus, to measure all of the $\mathcal{O}(N_{\text{bath}}^2)$ terms to a total measurement error $\varepsilon_m$, $N_s = \mathcal{O} \Big( N_{\text{bath}}^2/\varepsilon_m^{2} \Big)$ total samples are required. Substituting $\varepsilon_m \to \varepsilon_m/N_{\text{bath}}^2$  gives a total gate scaling of $\mathcal{O}(
\sqrt{\Upsilon} \varepsilon_s^{-1/2} \sqrt{t^3} N_{\text{bath}}^9\varepsilon_m^{-2})$.

Figure \ref{fig:trot} shows the ratio of the upper bound on the Trotter error from Ref.~\citenum{Kivlichan2020} to the actual Trotter error computed for our system in different parameter regimes for both the two and three-site systems. To obtain this figure, we compute the value of the factor $\Upsilon$ directly and utilize the 2-norm of the operators equal to the largest singular value.

\section{Conclusions} \label{sec:conclusions}
We have presented a hybrid quantum-classical approach for calculating time-domain Green's functions of fermionic models based on coupled-cluster methods. Applying this approach to the AIM, we built a rigorous but straightforward fermion-to-unitary mapping for the non-unitary CCGF exponential operators. We then combined the cluster amplitudes calculated using the classical coupled-cluster algorithm with measurements of time-evolution operators on a quantum device to extract the impurity Green's function in the real-time domain. On the quantum end, this approach replaces the need to prepare the ground state and can be easily generalized to many models. Our method has a T-gate count that scales as $\mathcal{O}([1-P_f]^{-1} \sqrt{\Upsilon} \varepsilon_s^{-1/2} \sqrt{t^3} N_{\text{bath}}^5 \varepsilon_m^{-2})$ with no state preparation, if we use the LCU measurement scheme outlined here with the Trotter (Givens rotation) time evolution procedure. 

We demonstrated the accuracy of our approach for the two- and three-level AIM models on a quantum simulator, where we obtained results in excellent agreement with the exact solutions. Further complexity analysis of the employed second-order symmetric time-evolution operator indicates that our method has comparable scaling of the gate counts (as a function of system size) in comparison with other state-of-the-art time evolution schemes. Our work provides a novel strategy for calculating the time-domain impurity Green's function using a controlled approximation for the ground state wave function. This approach could be beneficial for situations where we know little about the ground state of the system or when the system's spectral gap is small.

\bibliographystyle{plainnatnourl}
\bibliography{ref3}
\vspace{0.5cm}
\noindent
{\bf Acknowledgements}: We thank E.~F. Dumitrescu for useful discussions. T.~K. is supported by the U.~S. Department of Energy, Office of Science, Office of Workforce Development for Teachers and Scientists, Office of Science Graduate Student Research (SCGSR) program. The SCGSR program is administered by the Oak Ridge Institute for Science and Education (ORISE) for the DOE. ORISE is managed by ORAU under contract number DE‐SC0014664. All opinions expressed in this paper are the author’s and do not necessarily reflect the policies and views of DOE, ORAU, or ORISE. This work was performed in part at Oak Ridge National Laboratory, operated by UT-Battelle for the U.S. Department of Energy under Contract No. DE-AC05-00OR22725. B.~P. and K.~K. were supported by  the ``Embedding QC into Many-body Frameworks for Strongly Correlated  Molecular and Materials Systems'' project, which is funded by the U.~S. Department of Energy, Office of Science, Office of Basic Energy Sciences (BES), the Division of Chemical Sciences, Geosciences, and Biosciences. B.~P. also acknowledges the support of Laboratory Directed Research and Development (LDRD) program from PNNL. T.~K. and P.~L. were supported by the U.~S. Department of Energy, Office of Science, Office of Advanced Scientific Computing Research (ASCR) Quantum Algorithm Teams (QAT) and Quantum Computing Application Teams (QCATS) programs, under field work proposal numbers ERKJ333 and ERKJ347.\\

\noindent{\bf Author Contributions}: T.~K., B.~P., K.~K., and P.~L. designed the study, T.~K. and B.~P. collected data, and produced figures. S.~J. developed a classical exact diagionalization code for the AIM. B.~P., K.~K., and P.~L., supervised the research. All authors discussed the results and contributed to the final paper. \\

\noindent{\bf Competing Interests}: The authors declare no competing interests.\\

\noindent{\bf Data availability}: The data supporting this study are available at \url{https://doi.org/10.5281/zenodo.5920664}.\\

\noindent{\bf Code availability}: The code used to obtain the coupled-cluster quantities is available at \url{https://github.com/spec-org/gfcc}. The code used for the quantum-classical algorithm is available at \url{https://doi.org/10.5281/zenodo.5920664}. \\

\onecolumn\newpage
\appendix



\section{Finding Unitaries $\mathbf{W}$ and scalar coefficients $\{ \mu_i \}$.} \label{sec:ccstuff}

If index $p\in O$ denotes impurity site, then one option of $\mathbf{W}_p$ can be obtained through the Jordan-Wigner mapping of the transformed $c_p e^T|\Phi\rangle$. Here, we use the Jordan-Wigner transformation given by 
\begin{equation}
\begin{split}
& c_{j\downarrow}^\dagger = \frac{1}{2} \bigotimes_{k=0}^{j-1} Z_k (X_j - i Y_j) \\[1em]
& c_{j\uparrow}^\dagger = \frac{1}{2} \bigotimes_{k=0}^{N_{\text{bath}} + j} Z_k (X_{N_{\text{bath}} + j+ 1} - i Y_{N_{\text{bath}} + j+ 1}),
\end{split}
\end{equation}
where the first $N_\text{bath} + 1$ qubits hold the down occupation information for each site, and the second $N_\text{bath} + 1$ qubits hold the up occupation information for each site.
In the CCSD approximation, this can be expressed as
\begin{eqnarray*}
c_p e^T |\Phi \rangle &\approx&
c_p \Big( 1 + \sum_{ai} t_i^a c_a^\dagger c_i 
+ \sum_{a<b,i<j} \tilde{t}_{ij}^{ab} c_a^\dagger c_b^\dagger c_j c_i \Big) | \Phi \rangle \notag \\
&=& 
\Big( 1 + \sum_{a,i\neq p} t_i^a c_a^\dagger c_i 
+ \sum_{\substack{a<b,i<j\\i\neq p,j\neq p}} \tilde{t}_{ij}^{ab} c_a^\dagger c_b^\dagger c_j c_i \Big) c_p | \Phi \rangle \notag \\
&=& 
\Big( 1 + \sum_{a,i\neq p} t_i^a \big( c_a^\dagger + c_a \big) \big( c_i + c_i^\dagger \big)
+ \sum_{\substack{a<b,i<j\\i\neq p,j\neq p}} \tilde{t}_{ij}^{ab} \big( c_a^\dagger + c_a \big) \big( c_b^\dagger + c_b \big) \big( c_j + c_j^\dagger \big) \big( c_i + c_i^\dagger \big)\Big) \big( c_p + c_p^\dagger \big)  | \Phi \rangle \notag \\
&=&\Big( 
\tilde{X}_p + \sum_{a,i\neq p} t_i^a \tilde{X}_a\tilde{X}_i\tilde{X}_p
+ \sum_{\substack{a<b,i<j\\i\neq p,j\neq p}} \tilde{t}_{ij}^{ab} \tilde{X}_a \tilde{X}_b \tilde{X}_j \tilde{X}_i \tilde{X}_p \Big) | \Phi \rangle, \label{ap}
\end{eqnarray*}
where $\tilde{t}_{ij}^{ab} = t_{ij}^{ab} + t_i^a t_j^b - t_i^b t_j^a$. The $\mathbf{W}_p$ can then be chosen as $\big\{\tilde{X}_p, \{\tilde{X}_a\tilde{X}_i\tilde{X}_p\}_{a,i\neq p}, \{\tilde{X}_a\tilde{X}_b\tilde{X}_j\tilde{X}_i\tilde{X}_p\}_{a<b,i<j,i\neq p, j\neq p}\}$, and the scalars $\{\mu_i\}$ are the corresponding cluster amplitudes in (\ref{ap}).
$\langle \Phi |(1+\Lambda) e^{-T} c_q^\dagger$ can be expanded by the same set $\mathbf{W}_p$ with the corresponding scalars given by
\begin{eqnarray*}
\langle (1+\Lambda)e^{-T} c_q^\dagger c_p \rangle &=& 
\delta_{pq} \big( 1 - \sum_{i,a}\lambda_a^i t_i^a - \sum_{i<j,a<b} \lambda_{ab}^{ij} \tilde{t}_{ij}^{ab} \big) -  (1 - \delta_{pq}) \big( \sum_{a} \lambda_a^p t_q^a - \sum_{i,a<b} \lambda_{ab}^{pi} \tilde{t}_{qi}^{ab} \big), \notag \\
\langle (1+\Lambda)e^{-T} c_q^\dagger c_p c_a^\dagger c_i \rangle&=&
\delta_{pq} \big( \lambda_a^i -\sum_{j,b}\lambda_{ab}^{ij} t_{j}^b \big) \notag - (1-\delta_{pq}) \sum_{b} \lambda_{ab}^{ip} t_q^b, \notag \\
\langle (1+\Lambda)e^{-T} c_q^\dagger c_p c_a^\dagger c_b^\dagger c_j c_i \rangle 
&=&\delta_{pq} \lambda_{ab}^{ij}.
\end{eqnarray*}
Similarly, we can transform $c_q^\dagger e^T |\phi\rangle$,
\begin{eqnarray*}
c_q^\dagger e^T |\Phi \rangle &\approx &
c_q^\dagger \Big( 1 + \sum_{ai} t_i^a c_a^\dagger c_i 
+ \sum_{a<b,i<j} \tilde{t}_{ij}^{ab} c_a^\dagger c_b^\dagger c_j c_i \Big) | \Phi \rangle \notag \\
&=&
\Big( 1 + \sum_{a\neq q,i} t_i^a c_a^\dagger c_i 
+ \sum_{\substack{a<b,i<j\\a\neq q,b\neq q}} \tilde{t}_{ij}^{ab} c_a^\dagger c_b^\dagger c_j c_i \Big) c_q^\dagger | \Phi \rangle \notag \\
&=&\Big( 
\tilde{X}_q + \sum_{a\neq q,i} t_i^a \tilde{X}_a\tilde{X}_i\tilde{X}_q
+ \sum_{\substack{a<b,i<j\\a\neq q,b\neq q}}\tilde{X}_j \tilde{X}_i \tilde{X}_q \Big) | \Phi \rangle,
\end{eqnarray*}
to get another set of unitaries and corresponding scalars for greater part of the coupled cluster Green's function.

It is worth mentioning that the cluster amplitudes, $t$'s, in the present context are obtained from classical CCSD calculation on AIM, which numerically scales $~\mathcal{O}(N_s^2N_o^2)$ ($N_o$ denoting the number of occupied spin-orbitals and $N_s$ denoting the number of virtual spin-orbitals).

\end{document}